\documentclass[useAMS,usenatbib]{mn2e}
\usepackage{graphicx}

\title{New measurement of orbital and spin period evolution of the Accretion Disk Corona source 4U 1822-37}

\author[Chetana Jain, Biswajit Paul and Anjan Dutta]{Chetana Jain$^{1,2}$\thanks{E-mail: chetanajain11@gmail.com}, Biswajit Paul$^{1}$ and Anjan Dutta$^{2}$\\
$^{1}$Raman Research Institute, Sadashivnagar, C. V. Raman Avenue, Bangalore 560080, India\\
$^{2}$Department of Physics and Astrophysics, University of Delhi,  Delhi 110007, India}

\begin{document}
\pagerange{\pageref{firstpage}--\pageref{lastpage}} \pubyear{2010}
\maketitle
\label{firstpage}

\begin{abstract}

4U 1822$-$37 is a Low Mass X-ray Binary (LMXB) system with an Accretion Disk Corona. It is one 
of the very few LMXBs that shows narrow X-ray eclipses and small amplitude pulsations of the neutron star. The X-ray eclipse is an excellent reference for measurement of orbital evolution of the binary, and we have obtained 16 new mid-eclipse time measurements of this source during the last 13 years using X-ray observations made with the \emph{RXTE}-PCA, \emph{RXTE}-ASM, \emph{Swift}-XRT, \emph{XMM-Newton} and \emph{Chandra} observatories. These, along with the earlier known mid-eclipse times have been used to accurately determine the timescale for a change in the orbital period of 4U 1822$-$37. We have derived an orbital period $P_{orb}$ = 0.23210887(15) d, which is changing at the rate of $\dot{P}_{orb}$ = 1.3(3) $\times$ 10$^{-10}$ d d$^{-1}$ (at T$_{0}$ = MJD 45614). The timescale for a change in the orbital period is P$_{orb}$/$\dot{P}_{orb}$ of 4.9(1.1) $\times$ 10$^{6}$ yr. We also report the detection of 0.59290132(11) s (at T$_{0}$ = MJD 51975) X-ray pulsations from the source with a long term average $\dot{P}_{spin}$ of -2.481(4) $\times$ 10$^{-12}$ s s$^{-1}$, i.e., a spin-up time scale (P$_{spin}$/$\dot{P}_{spin}$) of 7578(13) yr. In view of these results, we have discussed various mechanisms that could be responsible for the orbital evolution in this system. Assuming the extreme case of conservative mass transfer, we have found that the measured $\dot{P}_{orb}$ requires a large mass transfer rate of (4.2 $-$ 5.2) $\times$ 10$^{-8}$ M$_{\odot}$ yr$^{-1}$ which together with the spin up rate implies a magnetic field strength in the range of (1$-$3) $\times$ 10$^{8}$ G. Using the long term \emph{RXTE}-ASM light curve, we have found that the X-ray intensity of the source has decreased over the last 13 years by $\sim$ 40$\%$ and there are long term fluctuations at time 
scales of about a year. In addition to the long term intensity variation, we have also observed significant variation in the intensity during the eclipse. Variation was also seen in the pulse profile, which could be due to changes in the accretion geometry. We briefly discuss the implications of these results on our understanding of the properties of the neutron star and the accretion geometry in this source. 
\end{abstract}
\begin{keywords}
accretion, accretion disks -- binaries: eclipsing, general -- stars:  individual (4U 1822$-$37) -- stars: neutron -- X-rays: stars
\end{keywords} 

\section{Introduction}
\label{Introduction}

Low mass X-ray binaries (LMXBs) are stellar systems that consist of a compact object, 
such as a neutron star or a black hole, accreting matter from a companion star by Roche lobe overflow. Mass transfer in these systems can be conservative or non-conservative. In most systems, the mass of the binary system and the total angular momentum remains constant, which is the case of conservative mass transfer. However, in some systems, a significant mass loss may occur and in such systems the mass transfer is called non-conservative. Mass loss from the binary system may occur through various processes such as, irradiative evaporation of the secondary star, jet emission from the compact star or emission of a wind from the accretion disk (Ruderman et al. 1989). Accretion may also be driven by the loss of orbital angular momentum through gravitational wave radiation, or magnetic braking (Hurley et al. 2002; Rappaport, Verbunt \& Joss 1983). The orbital period of X-ray binaries is expected to change due to redistribution of the angular momentum due to interaction between the components of the binary system. Measurement of the rate of change of the orbital period (i.e., orbital period derivative, $\dot{P}_{orb}$) of the binary system is therefore necessary in order to understand evolution of compact binary systems.

In some accretion powered X-ray binaries, repeated measurements of the 
orbital ephemeris have led to an accurate determination of the orbital period 
evolution. Pulse arrival time delay is one of the several techniques 
used to determine the evolution of the binary orbit. Using this technique, 
the orbital evolution timescales have been accurately determined in several X-ray binary 
systems (Her X-1: Deeter et al. 1991; Paul et al. 2004; Staubert, Klochkov $\&$ Wilms 2009, 
4U 1538$-$52: Baykal et al. 2006; Mukherjee et al. 2006, Cen X-3: Kelley et al. 1983; 
Paul et al. 2007, LMC X-4: Levine et al. 2000, SMC X-1: Levine et al. 1993, 
Wojdowski et al. 1998). Pulse folding and $\chi^{2}$ maximization with a varying 
orbital ephemeris have been successfully applied to LMC X-4 (Naik $\&$ 
Paul 2004) and SAX J1808.4$-$3658 (Jain et al. 2007) to determine the X-ray 
pulsations and the mid eclipse times. However, measurement of $\dot{P}_{orb}$ for a non-pulsing LMXB requires a stable fiducial point in the X-ray light curve. Such a measurement is difficult because most of the LMXBs do not exhibit sharp orbital features and even the pulsating objects often show variable pulse profiles.

In the absence of pulses from the compact object, eclipsing binary systems 
provide a good fiducial timing marker for precise determination of the 
orbital evolution. Parmar et al. (1986) and Wolff et al. (2009) applied 
the X-ray eclipse timing technique to determine the orbital evolution in 
the LMXB EXO 0748$-$676. Eclipse timings were also used to establish orbital 
evolution timescales in 4U 1822$-$37 (Hellier et al. 1990; Parmar et al. 2000) 
and 4U 1700$-$37 (Rubin et al. 1996).

In the case of non-eclipsing X-ray binaries, such as Cyg X-3, 
the orbital evolution has been measured using its stable orbital 
modulation light curve (Singh et al. 2002). Van der Klis et al. (1993) also 
studied the orbital evolution of the LMXB 4U 1820$-$303 by analyzing the 
modulation of the 685 s orbital light curve.

Amongst the above mentioned LMXBs, the orbital separation has been found to be 
increasing in 4U 1822$-$37 (Parmar et al. 2000), X2127$+$119 
(Homer $\&$ Charles 1998) and SAX J1808.4$-$3658 (Jain et al. 2007). The 
observed orbital evolution timescale of $\sim$10$^{6}$ yr in the case 
of 4U 1822$-$37 (Heinz $\&$ Nowak 2001) and X2127$+$119 (Homer $\&$ Charles 1998) 
is assumed to be due to short lived mass exchange episodes. An orbital evolution 
timescale of about 70 $\times 10^{6}$ yr has been determined in the accretion powered 
millisecond X-ray pulsar SAX J1808.4$-$3658 (Jain et al. 2007; Burderi et al. 
2009; Hartman et al. 2009). The timescale of orbital evolution was proposed to 
be due to a strong tidal interaction between the components of the binary. A 
decreasing orbital period has been detected in 4U 1820$-$303 (van der 
Klis et al. 1993) and Her X$-$1 (Deeter et al. 1991; Paul et al. 2004). 
However, the observed orbital period evolution is faster 
than that predicted by theoretical models of conservative mass transfer. 
In the case of EXO 0748$-$676, Wolff et al. (2009) found four distinct 
orbital period epochs in the last 20 years and attributed it to be due 
to magnetic cycling in the companion star.
 
4U 1822$-$37 is an LMXB with an orbital period of 5.57 hr (Hellier et al. 1990; Parmar et al. 2000). It is one of the few LMXBs that harbors an accretion powered X-ray pulsar (Jonker $\&$ van der Klis 2001). The source is surrounded by an accretion disc corona formed by evaporation of matter from the inner accretion disc by radiation pressure of the neutron star (White $\&$ Holt 1982). The light curve exhibits a narrow and a broad dip in the intensity. The narrow dip in the X-ray light curve is attributed to the partial eclipse of the corona by the companion star (White et al. 1981; White $\&$ Holt 1982; Mason $\&$ Cordova 1982; Hellier $\&$ Mason 1989, Hellier et al. 1992), whereas the broad dip is interpreted as resulting from occultation of the corona by a bulge on the outer edge of the accretion disc (White $\&$ Holt 1982; Hellier $\&$ Mason 1989). Being an eclipsing system, the orbital inclination is also known with a small uncertainty (Heinz $\&$ Nowak 2001; Jonker et al. 2003). An accurate ephemeris of 4U 1822$-$37 is also known from the eclipse timing (Parmar et al. 2000) and the size of the binary orbit is known from the pulse timing (Jonker $\&$ van der Klis 2001). It is thus an ideal system for determination of the masses of the stellar components. From the spectroscopic measurements of the binary system and by assuming a mass of 1.4 M$_{\odot}$ for the neutron star, Cowley et al. (2003) calculated the mass of the companion. A good estimate of the mass of the neutron star is also known from the K-correction of the radial velocity curves (Mu\~noz$-$Dariaz et al. 2005). 

We report here on a detailed timing analysis of the eclipsing X-ray binary pulsar, 4U 1822$-$37. We have determined 16 new mid-eclipse time measurements of the source with data obtained from instruments onboard \emph{RXTE}, \emph{Swift}, \emph{XMM-Newton} and \emph{Chandra} observatories. We have also measured the spin period from several newer \emph{RXTE} observations. Combining these new results with previously published measurements, we have derived new updated estimates of the orbital and spin parameters. Using these measurements, we have also determined the accretion luminosity and the magnetic field strength of the pulsar, which is crucial for this system. Even though the binary parameters have been reliably measured in the past, only rough estimates of the source intrinsic luminosity and the neutron star's magnetic field have been made by various authors, with inconsistent results. Assuming a distance of 2 kpc, Mason $\&$ Cordova (1982) had estimated an isotropic luminosity of $\sim$ 10$^{36}$ ergs s$^{-1}$. From the broad band spectrum analysis, Parmar et al. (2000) measured the 1$-$10 keV X-ray flux to be 5 $\times$ 10$^{-10}$ ergs cm$^{-2}$ s$^{-1}$. This implies an X-ray luminosity of 5.6 $\times$ 10$^{34}$ $(d/1 kpc)^{2}$ ergs s$^{-1}$. Assuming a magnetic field strength in the range (1$-$5) $\times$ 10$^{12}$ G, Jonker $\&$ van der Klis (2001) derived an intrinsic luminosity of (2$-$4) $\times$ 10$^{37}$ ergs s$^{-1}$ from the spin-up measurements. 

\section{Observations and analysis}

In this work, we have analyzed observations made with the Proportional Counter Array (PCA) and the All Sky Monitor (ASM) on board the Rossi X-ray Timing Explorer (\emph{RXTE}); the X-ray Telescope (XRT) on board \emph{Swift}; the European Photon Imaging Cameras (EPIC) - MOS instruments on board \emph{XMM-Newton}; and the Advanced CCD Imaging Spectrometer (ACIS) on board the \emph{Chandra} observatory. A log of the X-ray observations used for the present work is presented in Table 1. 

\begin{table}
\caption{Log of observations analyzed in the present work}
\centering
\begin{tabular}{c c c c c} 
\hline\hline
Instrument 	& Time range 	& Observation & Exposure \\
		& (MJD)		& IDs		& (ks)\\
\hline
\emph{RXTE}-ASM 	& 50088 - 54756		& 			 & 5096\\
\emph{RXTE}-PCA 	& 51018 	&30060-02-02-00		 & 18	\\
		&	51975 		&50048-01-01-05		 & 8	\\
		&	52094 		&50048-01-03-01 	 & 6	\\
		&	52095 		&50048-01-03-02		 & 8	\\
		&	52432 		&70037-01-01-00		 & 26	\\
		&	52489 		&70037-01-03-00		 & 21	\\
		&	52489 		&70037-01-03-01 	 & 7    \\
		&	52491 		&70037-01-04-00		 & 18	\\
		&	52503 		&70037-01-06-02 	 & 4   \\	
		&	52503 		&70037-01-06-04		 & 12	\\
		&	52519 		&70037-01-07-00		 & 24	\\
		&	52882 		&70037-01-11-00		 & 11	\\
		&	52883 		&70037-01-11-01		 & 13	\\
\emph{Chandra}-ACIS  	& 51779 - 51780 	& 671 			 & 40\\
\emph{XMM-Newton}-MOS & 51975 - 51976 	& 0111230101 		 & 54\\
\emph{Swift}-XRT  & 54317 - 54318 	& 00036691003		 & 17\\
\hline
\end{tabular}
\label{Table 1 }  
\end{table}

4U 1822-37 was monitored regularly by the \emph{RXTE}-ASM (Levine et al. 1996), 
which comprises three wide-field scanning shadow cameras (SSCs) which are mounted on a rotating boom. The SSC's are rotated in a sequence of $``$dwells$"$ with an exposure typically of 90 s, so that most of the sky can be covered in one day. The dwell data are also averaged for each day to yield a daily-average. The data used in the present analysis, covered the time between MJD 50088 to MJD 54756. The 1.5$-$12 keV long term ASM light curve was corrected for the earth motion using the tool \emph{earth2sun} of the HEAsoft analysis package, \emph{ftools} ver 6.5.1. The light curve, binned with a binsize of 50 days, is shown in Figure 1. It shows a gradual decrease in the X-ray intensity over the last 13 years, alongwith long 
term fluctuations over timescales of about an year. There could be many reasons for the observed decrease in the X-ray flux; decrease in mass accretion rate and a change in the structure of the corona (Bayless et al. 2010) being two possibilities.

To check whether the observed decrease in the source intensity is an instrumental effect, 
we also studied the light curve of the well known Crab Nebula. As shown in Figure 1, the intensity of Crab does not show a decreasing pattern over the same time period as in 4U 1822-37. The $RXTE$-ASM measurements of long term intensity variations are quite reliable and for some sources correlated variability has also been reported with different instruments (4U 1626-67: Jain et al. 2010). 

 \begin{figure}
 \centering
 \includegraphics[height=3.5in, angle=-90]{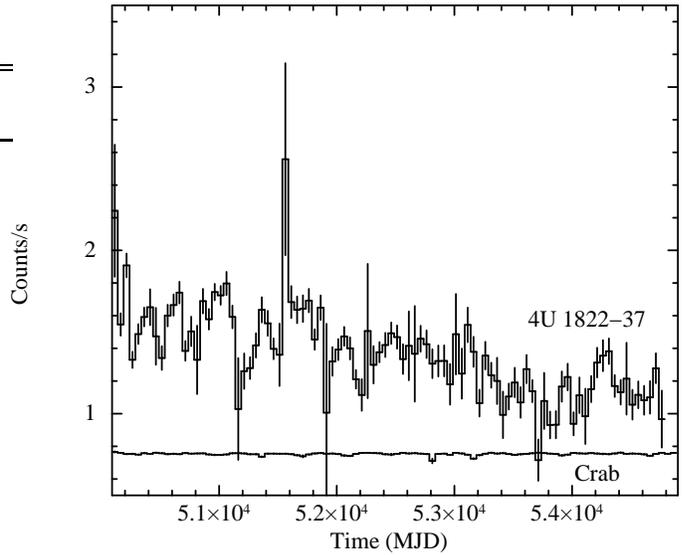}
 \caption{1.5$-$12 keV \emph{RXTE}-ASM light curve of 4U 1822-37 and Crab, 
binned with 50 d. An overall decrease in the intensity is clearly seen in 4U 1822-37, 
whereas the count rate is almost constant in case of Crab (Count rate for Crab has 
been scaled-down by a factor of 100).}
 \end{figure}
 
We have also analyzed the data taken with $RXTE$-PCA, consists of five xenon/methane proportional counter units (PCUs) and is sensitive in the energy range of 2$-$60 keV with an effective area of 1300$-$6500 cm$^{2}$, depending on the number of operating PCUs (Jahoda et al. 1996). Data for the eclipse timing analysis were chosen such that it covered the entire eclipse phase. The data were taken from the Standard-1 mode of PCA and the background count rate estimated using the \emph{runpcabackest} tool was subtracted from the light curves. The photon arrival times of the background subtracted light curve were then corrected 
for the solar system barycenter using the ftool \emph{fxbary}. To search for the pulsations, we have used the Event mode data with a time resolution of 16 $\mu$s.

4U 1822$-$37 was observed with \emph{XMM-Newton} (Jansen et al. 2001) for 54 ks on March 07, 2001 (obs ID - 0111230101). \emph{XMM-Newton} carries three X-ray mirrors and three focal plane instruments, a European Photon Imaging Camera (EPIC) - \emph{pn}, MOS1 and MOS2, each with a field of view of about 30$'$ $\times$ 30$'$. All the cameras (Struder et al. 2001; Turner et al. 2001) were operated in full frame mode with medium filter. The observation details are summarized in Table 1. The EPIC observation data files were processed using the \emph{XMM} - Science Analysis System (SAS version 8.0.0). For the present analysis, we have used data taken with the EPIC-MOS2 in the energy range 0.2$-$15 keV. The X-ray events were extracted from a circular region of radius 10$''$ centered on the position of the target in the EPIC-MOS image. The background X-ray events were extracted from a source-free 
circular region, with a radius of 20$''$. The background subtracted light curve was barycenter 
corrected using the SAS tool \emph{barycen}, using the JPL-DE405 ephemerides.

4U 1822-37 was observed with \emph{Chandra}-ACIS (Weisskopf et al. 2000) on August 23, 
2000 for an exposure time covering 2 full binary periods (Cottam et al. 2001). We used the Chandra Interactive Analysis of Observations (CIAO) software (ver. 4.0; CalDB ver.3.4.2) and standard Chandra analysis threads to reduce the data. No background flares were found, so all data were used for further analysis. For the present work, we used the energy range 0.3$–-$12.0 keV. Light curves were extracted from a circular region with a radius of 5$''$. Background events were obtained from an annular region with 
an inner (outer) radius of 15$''$ (30$''$). The background subtracted light curve was barycenter corrected using the CIAO-tool \emph{axbary}. It should be remarked that pile-up is an important phenomena which is inherent to CCD detectors, specially with instruments on board \emph{Chandra}. It is a major concern while measuring flux/spectra, especially when the source is bright. But there will be no effect on the determination of the mid-eclipse time. Pile-up can make the eclipse shallower but cannot change the mid-eclipse time. 

We have also analyzed data from the \emph{Swift} observatory (Gehrels et al. 2004). 
The scientific payload consists of a wide field instrument, the gamma ray Burst Alert 
Telescope (BAT; Barthelmy et al. 2005) and two co-aligned narrow field instruments: the X-ray 
Telescope (XRT, Burrows et al. 2005) operating in the 0.2-10 keV energy band and the Ultraviolet/Optical Telescope (UVOT, Roming et al. 2005). For the present work, the XRT data was processed with XRTDAS software data pipeline package (XRT-PIPELINE v.0.12.0). Calibrated and cleaned level 2 files were produced with the xrtpipeline task. We have used an energy range of 0.2$-$10 keV and all data were extracted in the Window Timing mode, with a total exposure time of 17.3 ks. X-ray events from within a rectangular region of width 6 and height 40 pixels were extracted for timing analysis. Background data was extracted from a neighbouring source free region of similar dimensions. We have applied \emph{earth2sun} correction to the background subtracted light curve which was produced with a timing resolution of 1 s. 

\subsection{Eclipse-timing analysis}

The light curves obtained from \emph{Chandra}, \emph{XMM-Newton}, \emph{Swift} and 
\emph{RXTE}-PCA observations were folded with the known orbital period of 5.5706 hr 
(Parmar et al. 2000). But since the $RXTE$-ASM dataset represents an average value over years of data, we folded the $RXTE$-ASM light curves with the best measurement of the orbital period derived from the actual $RXTE$-ASM data. We obtained an orbital period of 20054.27 s and 20054.24 s from the $RXTE$-ASM data spanning MJD 50088-52432 and MJD 52432-54756, respectively. The \emph{RXTE}-ASM light curve was therefore, folded in two segments (MJD 50088$-$52432 and MJD 52432$-$54756), while a single folded profile was obtained from each of the other observations. 

The folded light curves of the data obtained from \emph{RXTE}-ASM, \emph{Chandra}, 
\emph{XMM-Newton} and \emph{Swift} missions are shown in Figure 2. The \emph{RXTE}-PCA 
light curves are shown in Figure 3. It should be noticed that the eclipse 
morphology is changing. The eclipses seem to vary in depth and shape. The variable eclipse 
depth shows that the projected geometry of the accretion disk and corona is changing. It is also possible that a part of the change in the eclipse morphology in Figure 2, is due to the slightly different energy bands used and the differences in instrument efiiciency over the energy bands.

 \begin{figure}
 \centering
 \includegraphics[height=3.5in, width=5in, angle=-90]{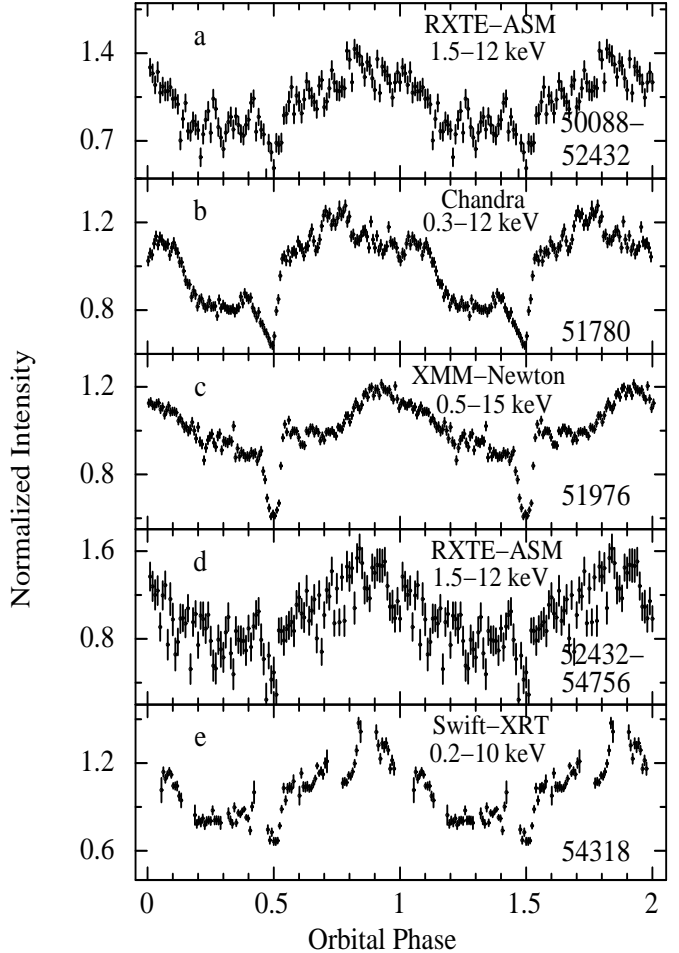}
 \caption{Folded light curves of 4U 1822-37 from the data obtained from \emph{RXTE}-ASM, \emph{Chandra}, 
\emph{XMM-Newton} and \emph{Swift}-XRT. The \emph{Chandra}, \emph{XMM-Newton} and \emph{Swift}-XRT 
light curves were folded with a period of 20054.3 s, while the $RXTE$-ASM light curves were 
folded with 20054.27 s (MJD 50088-52432) and 20054.24 s (MJD 52432-54756). The epoch of the mid eclipse time for each observation is given in each panel (in MJD). The Y-axis shows the 
intensity normalized by the average intensity and on the X-axis, two orbital cycles are shown for clarity.}
 \end{figure}

 \begin{figure}
 \centering
 \includegraphics[height=3.5in, width=6in, angle=-90]{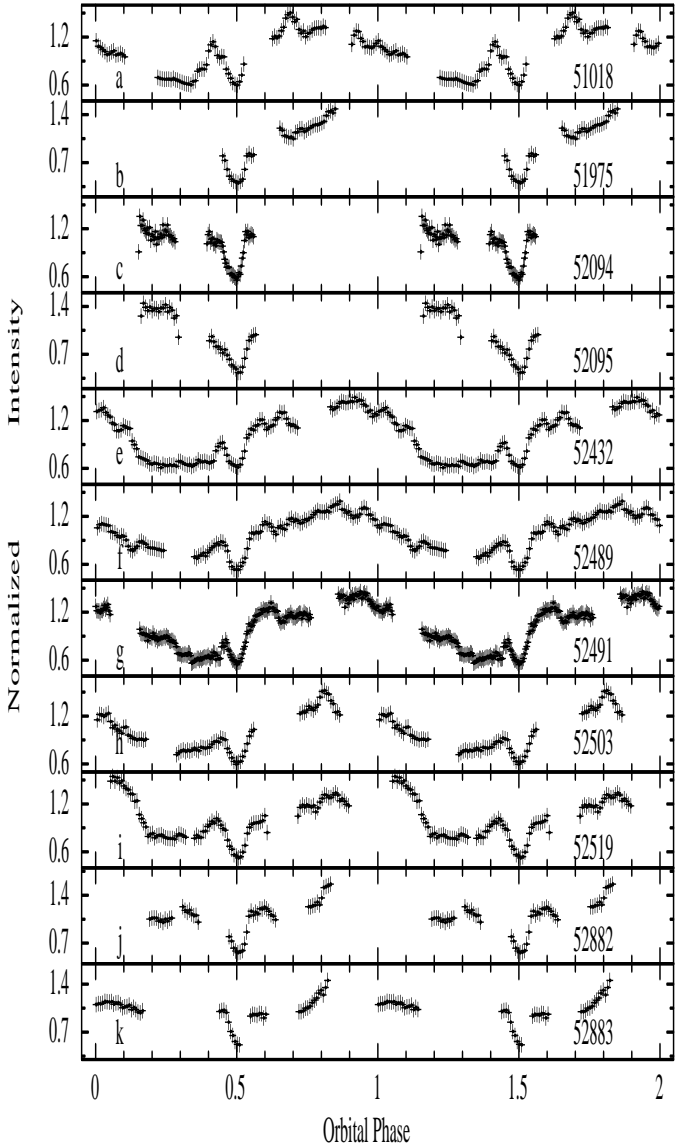}
 \caption{2$-$60 keV \emph{RXTE}-PCA light curves of 4U 1822-37, folded with a period of 20054.3 s. The epoch (in MJD) of each observation is given in the corresponding panel. The Y-axis shows the intensity normalized by the average intensity and on the X-axis, two orbital cycles are shown for clarity.}
 \end{figure}
 
\begin{table}
\caption{New X-ray mid-eclipse times of 4U 1822-37.}
\centering
\begin{tabular}{c l c c} 
\hline\hline
Cycle & MJD & Uncertainty (d) & Satellite\\
\hline
23281	&  51018.54400  &  0.00006	& \emph{RXTE}-PCA\\
24300	&  51255.0638  &  0.0018   	& \emph{RXTE}-ASM$\dag$\\
26562   &  51780.0954  &  0.0004 	& \emph{Chandra}\\
27406	&  51975.9968  &  0.0001 	& \emph{RXTE}-PCA\\
27407	&  51976.22941  &  0.00004	& \emph{XMM-Newton}\\
27918	&  52094.83802  &  0.00007	& \emph{RXTE}-PCA\\
27922	&  52095.76593  &  0.00006 	& \emph{RXTE}-PCA\\
29374	&  52432.78961  &  0.00005	& \emph{RXTE}-PCA\\
29618	&  52489.42192  &  0.00005	& \emph{RXTE}-PCA\\
29628	&  52491.74614  &  0.00007	& \emph{RXTE}-PCA\\
29678	&  52503.34773  &  0.00005	& \emph{RXTE}-PCA\\
29747	&  52519.36724  &  0.00006	& \emph{RXTE}-PCA\\
31311	&  52882.38623  &  0.00006 	& \emph{RXTE}-PCA\\
31315	&  52883.3151  &  0.0001 	& \emph{RXTE}-PCA\\
34378	&  53594.2680  &  0.0020 	& \emph{RXTE}-ASM$\dag$\\
37498	&  54318.4506  &  0.0004 	& \emph{Swift}-XRT\\
\hline
\end{tabular}
$\dag$The mid eclipse times represent the time span MJD 50088-52432 and 52432-54756.
\label{Table 2}  
\end{table}

The light curves show clear signs of orbital modulation (i.e. partial eclipse and a sinusoidal modulation) with an orbital period of 5.57 hr. A model consisting of a Gaussian and a constant was fit to the eclipse interval (0.45$-$0.55 orbital phase) in each folded light curve (as in Parmar et al. 2000). The reduced $\chi^{2}$ of the fits were in the range 0.5$-$4 for 13 d.o.f. Figure 4 shows the eclipse interval (0.45$-$0.55 orbital phase) for one of the folded \emph{RXTE}-PCA light curves (ObsId 70037-01-03-00). Gaussian and a constant model was fit to this interval. Solid line in the top panel of Figure 4 shows the best fit model and the bottom panel shows the residuals of the fit. 

 \begin{figure}
 \centering
 \includegraphics[height=3.5in, width=5.0in, angle=-90]{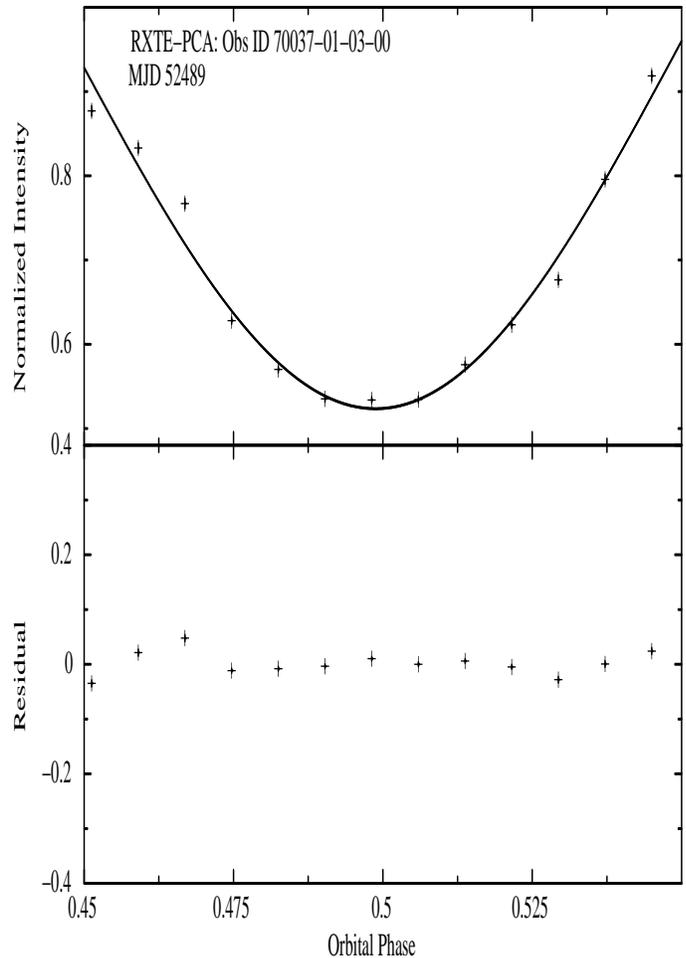}
 \caption{The expanded view of the eclipse interval (0.45$-$0.55 orbital phase) of one of the 
folded \emph{RXTE}-PCA light curve of 4U 1822-37 (ObsId 70037-01-03-00; Figure3(f)). A model 
consisting of a Gaussian plus a constant was fit to this interval. The solid line in the top panel shows the best fit model (with a reduced $\chi^{2}$ of 1.2 for 13 d.o.f.) and the bottom panel shows the residuals of the fit.}
 \end{figure}
 
The arrival time of the eclipse which occurred closest to the mid time of the observation was taken for further analysis. The new mid eclipse time measurements along with 1$\sigma$ uncertainties are given in Table 2. The orbit number (cycle) is with respect to the first reported mid-eclipse time (Hellier $\&$ Smale 1994). The newly determined arrival times were combined with the earlier known values and fitted with a quadratic model. We obtained a $\chi^{2}$ of 432 for 35 d.o.f. However, it should be remarked that the uncertainties in the newly determined mid eclipse time measurements are too small to give a reliable estimate. Therefore, we rescaled the errors in the individual measurements (by multiplying the individual errors by the square root of the above mentioned reduced $\chi^{2}$), in order to compare the results with the earlier known estimates of the orbital parameters (Hellier $\&$ Smale 1994; Parmar et al. 2000). The 16 new arrival times were combined with the earlier observations and were again fitted with a quadratic model to obtain an updated ephemeris. The best fit model gave an orbital period (P$_{orb}$) of 0.23210887(15) d and a period derivative ($\dot{P_{orb}}$) of 1.3(3) $\times$ 10$^{-10}$ d d$^{-1}$ (at T$_{0}$ = MJD 45614) with 
a $\chi^{2}$ of 36.2 for 35 d.o.f. This imply an orbital evolution timescale 
of 4.9(1.1) $\times$ 10$^{6}$ yr. The resulting best fit orbital parameters are given in 
Table 3. We subtracted the best fit linear component from the ephemeris history and the 
residuals are plotted in Figure 5. The derived values of the rate of change in orbital 
period and the timescale of orbital evolution are consistent, within measurement errors, 
with those obtained from timing of the eclipses in the optical and UV data (Bayless et al. 2010).

 \begin{figure}
 \centering
 \includegraphics[height=3.5in, width=4.0in, angle=-90]{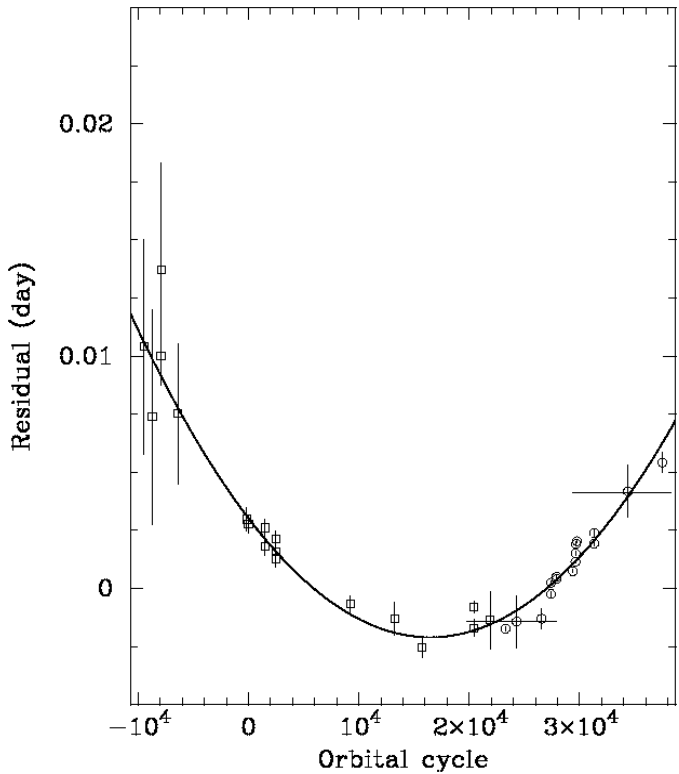}
 \caption{The mid eclipse time residuals of 4U 1822-37 have been plotted as a function of the 
orbital cycle, relative to the best fit linear ephemeris (orbital period of 0.23210854 d). The solid curve is the best fit of a quadratic function having a $\chi^{2}$ of 36.2 for 35 d.o.f. The eclipse times are tabulated in Table 2. The curvature of the locus of the residuals is a measure of the orbital period derivative of the binary system. The square boxes are the values known from previous measurements. The newly determined mid eclipse times are shown 
with $``\circ"$. The two horizontal bars indicate the time span of the \emph{RXTE}-ASM data.}
 \end{figure}

\begin{table*}
\caption{Orbital parameters of 4U 1822-37. The numbers in the bracket indicate the 1$\sigma$ uncertainty.}
\begin{center}
\begin{tabular}{c c c c}
\hline\hline
Parameter & Hellier $\&$ Smale  & Parmar \emph{et al.} & Best fit value from \\
 & (1994) & (2000) & the present analysis\\
\hline
T$_{0}$ (MJD) & 45614.80962(15) & 45614.80964(15) & 45614.80951(12)\\
P$_{orb}$ (d)  & 0.232108788(64) & 0.232108785(50) & 0.23210887(15) \\
c (10$^{-11}$ days)$\dag$ & 2.37 $\pm$ 0.56 & 2.06(23) & 1.53(40)\\
P$_{orb}$/$\dot{P}_{orb}$ (10$^{6}$ yr) & 3.1 $\pm$ 0.7 & 3.6 & 4.9(1.1)\\
\hline
\end{tabular}
\end{center}
$\dag${Quadratic Ephemeris: JD = T$_{0}$ + N$\times$P$_{orb}$ + N$^{2}$ $\times$ c}
\label{Table 3}
\end{table*}

\subsection{Pulse-timing analysis}

We have performed a pulsation analysis to determine the spin period of 
the neutron star and the pulse period evolution. The light curves were corrected for 
the orbital motion using the long term orbital solution obtained from the eclipse timing 
technique described above. Figure 6 shows the spin period history of the neutron star.
The pulse period was found to be continuously decreasing with time at an average 
rate ($\dot{P_{spin}}$) of -2.481(4) $\times$ 10$^{-12}$ s s$^{-1}$, indicating a spin-up 
timescale of 7578(13) yr. From the observations made in 1996 and 1998, Jonker $\&$ van der 
Klis (2001) had calculated an average spin-up ($\dot{P_{spin}}$/P$_{spin}$) 
of -1.52 $\times$ 10$^{-4}$ yr$^{-1}$. The pulse profiles obtained after correcting for the orbital motion, were normalized by the average intensity and are shown in Figure 7. The value of the spin period, determined by Jonker $\&$ van der Klis (2001) are also marked for completeness. A considerable variation was seen in the pulse profile. The pulse profile exhibited a non-sinusoidal profile at T$_{0}$ = MJD 51975; a relatively broad maximum at T$_{0}$ = MJD 52432; a sharper profile at around T$_{0}$ = MJD 52491 and a sinusoidal variation around T$_{0}$ = MJD 52519. However, it is difficult to quantify the observed variation in the pulse shape at this point.  

We also tried to obtain an independent measurement of orbital evolution 
using the technique of pulse folding and $\chi^{2}$ maximization (Naik $\&$ Paul 2004; 
Jain et al. 2007). But it should be remarked that in this source, the light travel time 
across the orbit ($a_{x} \sin i$) is only a factor of two larger than the spin period. 
Moreover, in case of pulse timing  technique, the pulse profile is assumed to be invariant. 
But even a small orbital phase dependence of the pulse shape, for example caused by varying 
absorption can lead to systematic errors in measurement of the orbital parameters. Therefore, 
though this analysis has proved to be successful for other LMXBs, we could not find a very 
accurate measurement of the orbital parameters by this technique. 

\begin{table}
\caption{Mid-eclipse times and the corresponding pulse period.}
\begin{center}
\begin{tabular}{| c | c |}
\hline\hline
\multicolumn{1}{|c|}{Mid-eclipse times (MJD)} & \multicolumn{1}{l|}{Spin period (s)}\\
\cline{1-2}
51975.9968(1)	&	0.59290132(11)\\
52094.83802(7)	&	0.59286109(8)\\
52095.76593(6)	&	0.59286421(12)\\
52432.78961(5)	&	0.5927922(13) \\
52489.42192(5)	&	0.5927790(06) \\
52491.74614(7)	&	0.5927795(11) \\
52503.34773(5)	&	0.5927737(10) \\
52519.36724(6)	&	0.5927721(08) \\
52882.38623(6)	&	0.5926793(15) \\
52883.3151(1)	&	0.5926852(21) \\
\hline
\end{tabular}
\end{center}
\label{Table 4}
\end{table}

 \begin{figure}
 \centering
 \includegraphics[height=3.5in, width=3.5in, angle=-90]{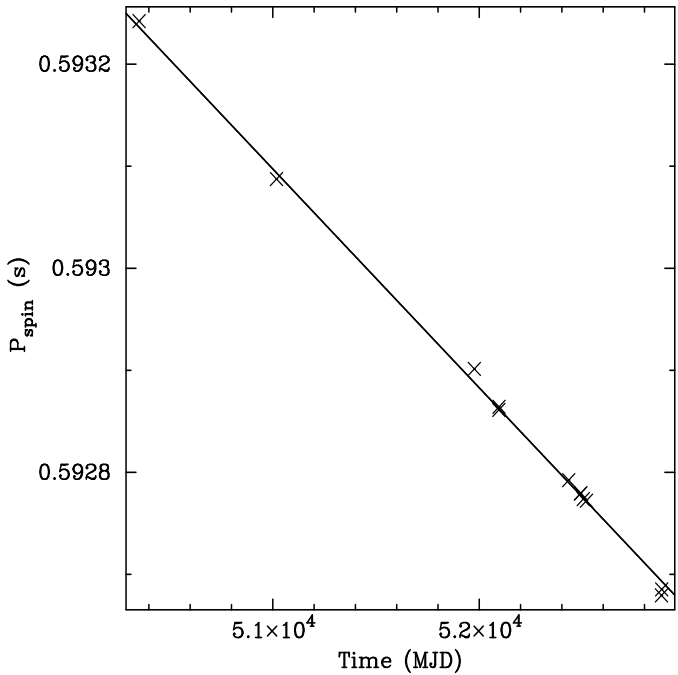}
 \caption{The variation of spin period (P$_{spin}$) of the neutron star with time. The first two points indicate the spin period determined by Jonker $\&$ van der Klis (2001). The straight line represents the best fitted linear curve, with a $\chi^{2}$ of 13.7 for 12 d.o.f.}
 \end{figure}

 \begin{figure}
 \centering
 \includegraphics[height=3.5in, width=5.0in, angle=-90]{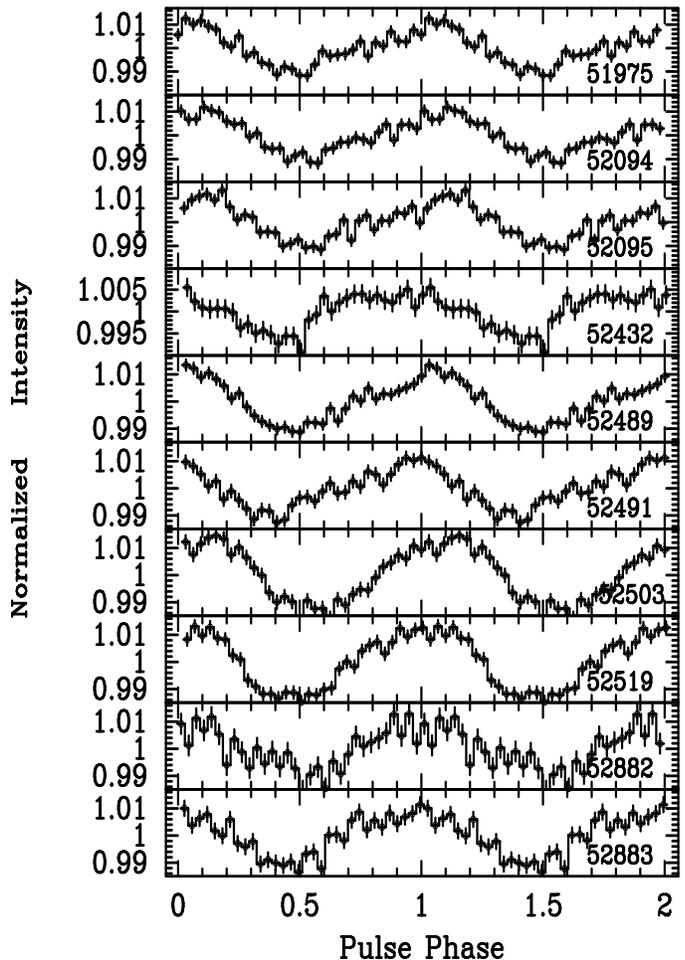}
 \caption{Normalized X-ray pulse profiles of 4U 1822-37, obtained from observations 
with \emph{RXTE}-PCA. We have divided the pulse period into 32 bins. The epoch (in MJD) of each observation is mentioned in the corresponding panel.}
 \end{figure}
 
\section{Discussions}

We have performed a detailed timing analysis of the low mass X-ray binary pulsar 4U 1822$-$37 using the archival X-ray data from several X-ray observatories. Observations used in the present work covered a time span of 13 years and more than 14,000 binary orbits. Using the 16 new accurately measured mid-eclipse times, we have obtained an orbital period of 0.23210887(15) d with a significant orbital period derivative of 1.3(3) $\times$ 10$^{-10}$ d d$^{-1}$ (at T$_{0}$ = MJD 45614). It indicates an orbital evolution timescale 
P$_{orb}$/$\dot{P}_{orb}$ = 4.9(1.1) Myr. The orbital and spin parameters were 
also measured by correcting the light curves for the binary motion of the pulsar and then optimizing the pulse detection. However, the results from the pulsation analysis did not improve the orbital evolution measurements. 

\subsection{Orbital Evolution}
X-ray binaries can evolve by various mechanisms such as mass transfer within the system due to 
Roche lobe overflow, tidal interaction between the components of the binary system, gravitational wave radiation, magnetic braking, and X-ray irradiated wind outflow (As mentioned before in Section 1). Orbital evolution has been measured in some other low magnetic field LMXBs, such as 4U 1820$-$30 (van der Klis et al. 1993), EXO 0748$-$676 (Wolff et al. 2008) and SAX J1808.4$-$3658 (Jain et al. 2007), and several models have been proposed to explain the orbital period evolution in them. In the case of 4U 1822-37, the reason behind a high rate of orbital evolution is not known. The measured rate of change of orbital period in 4U 1822$-$37 is much greater than that expected due to gravitational wave radiation (Verbunt 1993). We also note that the timescales of orbital evolution due to tidal interaction between the components of the binary system ranges from a few Myr in HMXBs to about 10$^{10}$ years in LMXBs (Applegate $\&$ Shaham 1994). Conservative mass transfer, mass loss from the binary system due to an X-ray irradiated wind outflow and magnetic cycling in one of the binary components are the other possibilities. We first examine the possibility and 
consequences of the case of a conservative mass transfer in this system, followed by the case 
of an X-ray irradiated wind outflow and magnetic cycling in the companion star.

\subsubsection{Conservative mass transfer}
In the case of conservative mass transfer, the mass transfer rate from the companion star 
equals the accretion rate onto the neutron star. We have estimated the mass accretion rate 
using the best known estimate of the mass of the neutron star (M$_{ns}$) and the companion 
star (M$_{c}$) (Mu\~noz-Dariaz et al. 2005) in this system:
\begin{eqnarray}
1.61 M_{\odot} \leqslant M_{ns} \leqslant 2.32 M_{\odot} \\
0.44 M_{\odot} \leqslant M_{c}  \leqslant 0.56 M_{\odot} 
\end{eqnarray} 
and a mass ratio q $\left( = \frac {M_{c}} {M_{ns}} \right)$ in the range:
\begin{eqnarray}
0.24 \leqslant q \leqslant 0.27
\end{eqnarray} 
In the case of compact binaries, the orbital angular momentum (J) is given by (King 1988):
\begin{eqnarray}
J = M_{c}M_{ns} \left( \frac{Ga} {M}\right)^{\frac{1}{2}}
\end{eqnarray} 
where, 
\begin{tabbing}
M \= = Total mass of the binary system = $M_{c}$ + $M_{ns}$\\
G \> = Gravitational constant\\
a \> = Binary orbital separation.
\end{tabbing}

For conservative mass transfer, all the mass lost from the companion is accreted by the 
compact object ($\dot{M_{ns}} + \dot{M_{c}} = 0$) and the orbital angular momentum is 
conserved ($\dot{J}$ = 0). In such a case, the mass accretion rate ($\dot{M_{ns}}$) is 
related to the orbital period ($P_{orb}$) as:
\begin{eqnarray}
\dot{M_{ns}} = \frac{1}{3} M_{c} \frac{\dot{P}_{orb}}{P_{orb}} \left( 1-\frac{M_{c}} {M_{ns}}\right)^{-1}
\end{eqnarray}

Since the observed $\frac {\dot{P}_{orb}} {P_{orb}}$ is 2.10 $\times$ 10$^{-7}$ yr$^{-1}$, therefore, we get a mass accretion rate of (4.2$-$5.2) $\times$ 10$^{-8}$ M$_{\odot}$ yr$^{-1}$ and a mass transfer timescale of $\sim$ 10$^{7}$ yr. Assuming an accretion efficiency ($\eta$) of 10$\%$, this implies an accretion luminosity L$_{acc}$ (= $\eta \dot{m} c^{2}/2$) in the range (1.19$-$1.48) $\times$ 10$^{38}$ ergs s$^{-1}$. 

For a neutron star with a magnetic moment $\mu$ ($\sim$ B$r^{3}$), the spin frequency 
derivative $\dot{\nu}$ is related with the mass accretion rate as (Frank et al. 2002): 
 \begin{eqnarray}
 \dot{\nu} \simeq 0.47 \times 10^{-12} m_{1}^{3/7} \dot{m_{16}}^{6/7} \mu_{30}^{2/7} I_{45}^{-1} \textrm{Hz s}^{-1}
 \end{eqnarray}
 where, 
 \begin{tabbing}
 \=m$_{1}$ \= = mass (m) of the neutron star in units of 1 M$_{\odot}$\\
 \>$\dot{m_{16}}$ \> = accretion rate in units of 10$^{16}$ g s$^{-1}$\\
 \>$\mu_{30}$ \> = magnetic moment in units of 10$^{30}$ G cm$^{3}$\\
 \>I$_{45}$ \> = the moment of inertia of the accreting star in units of \\
10$^{45}$ g cm$^{2}$. I$_{45}\sim$1 for a neutron star.
\end{tabbing}

Using data obtained from \emph{RXTE}-PCA, we have detected 0.59290132(11) s (at T$_{0}$ = MJD 51975) X-ray pulsations with an average spin-up rate of -2.481 $\times$ 10$^{-12}$ s s$^{-1}$. These values imply a magnetic field strength of (1$-$3) $\times$ 10$^{8}$ G. 

The following caveats apply to the case of conservative mass transfer:

\begin{itemize}
\item It requires a large mass transfer which corresponds to luminosity near the 
Eddington rate. Considering the uncertainties in some of the parameters used above, a near-Eddington accretion luminosity is not inconceivable, especially because this source has a corona surrounding it (White $\&$ Holt 1982; Heinz $\&$ Nowak 2001; Bayless et al. 2010). A comparison of the L$_{x}$/L$_{opt}$ ratio of this system with other LMXBs also suggest that the true X-ray luminosity of the central X-ray source is probably significantly higher. See Bayless et al. (2010) for more discussion on this. For example, in SS433 the mass transfer rate is believed to be much higher than the Eddington rate, that results in accumulation of material around the compact object which blocks its X-ray emission (Begelman, King, $\&$ Pringle 2006; Clark, Barnes, $\&$ Charles, 2007).

\item It requires the neutron star to have a low magnetic field strength of (1$-$3) $\times$ 10$^{8}$ G while from the high energy cutoff in the X-ray spectrum (Parmar et al. 2000), Jonker $\&$ van der Klis (2001) determined a magnetic field strength of $\sim$(1$-$5)$\times$10$^{12}$G. It should be noted that the coronal X-rays dominate the X-ray spectrum of this source and the pulsed X-rays contribute to only a few percent of the total X-ray emission. Thus the X-ray spectral shape is unlikely to be a reliable indicator of the magnetic field strength of the compact star. 

\item
In accreting neutron stars, we expect pulsations only if the compact object possesses a magnetic field strong enough to disrupt the inner regions of the accretion disc and channel the accretion flow onto the polar caps. In other words, accretion onto the neutron star is controlled by the magnetic field if the magnetospheric radius is larger than the stellar radius. In case of 4U 1822-37, the observed values of the mass transfer rate and the magnetic moment, indicate a magnetospheric radius of ~2 $\times$ 10$^{6}$ cm, which is of the same order as the neutron star radius. Therefore, it is not certain whether the conservative mass transfer is indeed responsible for the observed changes in the orbital period. However, several low mass X-ray binaries, such as SAX J1808.4-3658, do show pulsations at similar luminosity level even though they have a low surface magnetic field of $(1{-}5) \times 10^8$ G (Di Salvo $\&$ Burderi 2003).

\item For the accretion luminosity, spin period and magnetic field strength mentioned above for a conservative mass transfer case, the neutron star is far from spin equilibrium. Detection of such a system is a novelty and unlikely, unless the current X-ray state is a long lived transient phase. Here one may note that in the recent years, many LMXBs have been found which spend a small fraction of the time in transient high states, for example, the millisecond accreting pulsars.

\end{itemize}

\subsubsection{X-ray irradiated wind outflow}
Mass loss from the binary system can occur in the form of X-ray irradiated wind outflow. 
In case of LMXB 4U 1822-37, there is no signature of wind outflow in the form of absorption 
lines in high resolution X-ray spectrum (Cottam et al. 2001). Recently, Bayless 
et al. (2010) reported a broad C IV emission line in the UV spectrum of 4U 1822-37, indicating 
a strong disk wind outflow. They have derived a wind outflow velocity of 4000 km s$^{-1}$, based on a measured width of 45 $A^{\circ}$ of the C IV emission line. However, these observations are insufficient to estimate the total mass outflow rate. 

\subsubsection{Magnetic cycling in the Companion star}

Secular changes such as magnetic cycles in the secondary star (Hellier et al. 1990) is also a 
possible mechanism responsible for the observed orbital evolution in 4U 1822-37. 
However, the spectral type and the evolutionary history of the companion star is 
unknown (Mu\~noz$-$Dariaz et al. 2005). 

\section{Conclusions}
We have presented new and more accurate measurement of orbital evolution of the LMXB 4U 1822--37 and a longer time base for measurement of its spin evolution. Considering the possibility of a large intrinsic L$_{x}$ and some evidence of mass outflow, we conclude that the orbital evolution in this system is complex including the effects of a large mass transfer rate and X-ray irradiated wind outflow. In this scenario, the magnetic field strength of the neutron star is probably in between the same in typical low mass X-ray binaries and high mass X-ray binaries.

\section{Acknowledgement}
We are grateful to the anonymous referee for some useful comments in improving the manuscript. 
This research has made use of the data obtained through the High Energy Astrophysics Science 
Archive Research Center Online Service, provided by the NASA/Goddard Space Flight Center. In 
particular, we thank the \emph{RXTE}-ASM teams at MIT and at the \emph{RXTE}-SOF 
and GOF at NASA's GSFC for provision of the ASM data.

\label{lastpage}
\end {document}